\documentclass[aps,pra,10pt,twocolumn,superscriptaddress,floatfix,nofootinbib,showpacs,longbibliography]{revtex4-2}

\usepackage{physics}
\usepackage{amssymb}
\usepackage{graphicx}
\usepackage{lipsum}
\usepackage{adjustbox}
\usepackage{graphicx} % Required for inserting images
\usepackage{xcolor}
\usepackage{relsize}
\usepackage[colorlinks=true,linkcolor=blue,urlcolor=magenta,citecolor=red]{hyperref} 
\usepackage{amsfonts} 
\usepackage{comment}

\graphicspath{ {./figures/} }

\begin{document}

%================================================================================

\title{Long distance quantum illumination and ranging using polarization\\  entangled photon pairs in a lossy environment}

\author{Sujai Matta}
\altaffiliation{These authors contributed equally to this work}
	\affiliation{Quantum Optics \& Quantum Information, Department of Electronic Systems Engineering, Indian Institute of Science, Bengaluru 560012, India.}
\author{Soumya Asokan}
\altaffiliation{These authors contributed equally to this work}
	\affiliation{Quantum Optics \& Quantum Information, Department of Electronic Systems Engineering, Indian Institute of Science, Bengaluru 560012, India.}
\author{Sanchari Chakraborti}
	\affiliation{Quantum Optics \& Quantum Information, Department of Electronic Systems Engineering, Indian Institute of Science, Bengaluru 560012, India.}
\author{Mayank Joshi}
	\affiliation{Quantum Optics \& Quantum Information, Department of Electronic Systems Engineering, Indian Institute of Science, Bengaluru 560012, India.}
\author{Rahul Dalal}
	\affiliation{Quantum Optics \& Quantum Information, Department of Electronic Systems Engineering, Indian Institute of Science, Bengaluru 560012, India.}
\author{C. M. Chandrashekar}
	\email{chandracm@iisc.ac.in}
	\affiliation{Quantum Optics \& Quantum Information, Department of Electronic Systems Engineering, Indian Institute of Science, Bengaluru 560012, India.}
	% \affiliation{Homi Bhabha National Institute, Training School Complex, Anushakti Nagar, Mumbai 400094, India}

%================================================================================

%================================================================================

\begin{abstract}
\noindent{Using polarization entangled photon pairs, we demonstrate a robust scheme for quantum illumination and ranging in a lossy environment.  Entangled photon pairs are generated in a Sagnac interferometer configuration, yielding high-visibility two-photon polarization entanglement with a measured CHSH parameter of $S =2.802\pm0.002$. One of the photons from the entangled pair is retained as idler and the other one is directed into either of the two paths, namely reference and probe, of which probe is sent toward a distant object through a lossy  free-space channel, and the reflected photons are collected after round-trip free-space propagation over distances approaching $1$ km. Remarkably, strong correlations are observed with CHSH values $S >2.6$ even when only a few tens of probe photons are returned, confirming the robustness of polarization entanglement under long-distance free-space propagation. This work reports the robustness of encoding photons in different basis before it is sent towards the object and recovery of polarization entanglement even after a kilometer-scale scattering from the objects, establishing a practical foundation for scalable quantum-assisted object detection and ranging.}
\end{abstract}

%================================================================================

\maketitle

%================================================================================

\noindent
\textit{Introduction:--} Quantum entanglement lies at the heart of quantum mechanics and serves as the essential resource for several foundational aspects as well as for many quantum information processing protocols \cite{bouwmeester1997, jennewein2000, patel2016, flamini2019}. However, entangled quantum states are highly sensitive to decoherence, losses and environmental noise, which limits their practical deployment in long-distance quantum networks and field-scale applications. Despite these challenges, free-space quantum communication and quantum sensing systems using entangled state of light have made remarkable progress in the recent years, showing that quantum correlations can be distributed over increasingly long distances \cite{bulla2023, sajeed2021, zhuang2025, zhang2015}. Quantum sensing not only targets higher sensitivity in physical measurements but also extends to the standoff detection of remote objects under noisy conditions \cite{pirandola2018, england2019, pirandola2021}. Quantum illumination (QI) exploits non-classical correlations between entangled pair of photons to enhance object detection efficiency and ranging performance in the presence of noise and loss, offering advantages over the existing classical optical schemes \cite{shapiro2020, karsa2024, ushadevi2009, shapiro2009, guha2009, zhao2025}. \\

\noindent
The first theoretical formulation of QI proposed a scheme for locating a low-reflectivity object embedded in a noisy background using entangled photons \cite{Lloyd2008}.
In this approach, the signal photon is transmitted toward an object, while its entangled partner (the idler) is retained at the receiver. A joint correlation measurement, performed at the receiving end between the idler photon and the received signal photon that returns after reflection from the distant object, enables superior discrimination between the presence or absence of an object at a distance, even when the environmental decoherence has largely destroyed the initial entanglement. A comprehensive theoretical treatment of QI using Gaussian entangled states demonstrated that, despite the loss of entanglement during propagation, a measurable quantum advantage remains over any classical illumination methods. Their analysis, based on quantum Chernoff bounds, showed that such systems can achieve up to a $6$ dB enhancement in the error-exponent relative to classical schemes, thereby establishing the resilience of quantum correlations for sensing tasks in noisy and lossy environments \cite{tan2008}. Later, experimental demonstrations of QI were performed using entangled photon pairs produced from spontaneous parametric down-conversion\,(SPDC) process, verifying efficient object detection under noise with non-classical correlation measurement between the signal and idler photons as compared to using classical light sources \cite{lopaeva2013, lopaeva2014}. Over the years, several studies have been reported demonstrating quantum-enhanced illumination, highlighting its potential to surpass classical detection strategies by exploiting non-classical correlations \cite{chang2019, xu2021, zhang2020, kuniyil2022}.\\

\noindent
In addition to engineering entanglement of photon pairs in polarization degree of freedom, they can also exhibit entanglement in different degrees of freedom, such as their path and polarization. These systems are shown to be fundamental to the continued progress and implementation of next-generation quantum technologies\, \cite{solntsev2017, casper2020, casper2022, abiuso2022, shimizu2025}. Extending the Hilbert space using different degrees of freedom also offers a viable physical realization of quantum bits \cite{qc1, qc2}. Theoretical studies have demonstrated that the use of hyper-entangled probe states can further enhance the efficiency of QI \cite{prabhu2021, sairam2022}.  Recent experimental studies on QI have employed three distinct paths for the correlated pair of photons - with only one probing the target, to further suppress the background noise that interferes with the signal \cite{shafi2023, kanad2024}. Employing polarization-path entanglement and polarization-entanglement, the authors showed significant suppression of background noise while detecting an object through the evaluation of the CHSH-Bell parameter ($S$-value) to quantify the quantum correlations between the pair of photons upon receiving the probe photon after reflection from the object. Even when the reflected signal was strongly masked by the background thermal noise, with signal-to-noise\,(SNR) ratios being $0.03$~\cite{shafi2023} and $0.002$~\cite{kanad2024}, values of $S > 2$ were recorded confirming the presence of the object. Extending this approach, we present a novel QI protocol employing polarization-entangled photon pairs and demonstrate its implementation in a real-world field experiment, achieving kilometer-scale object detection under realistic atmospheric conditions involving huge loss.\\

\begin{figure}[h]
\centering
\includegraphics[width=0.49\textwidth]{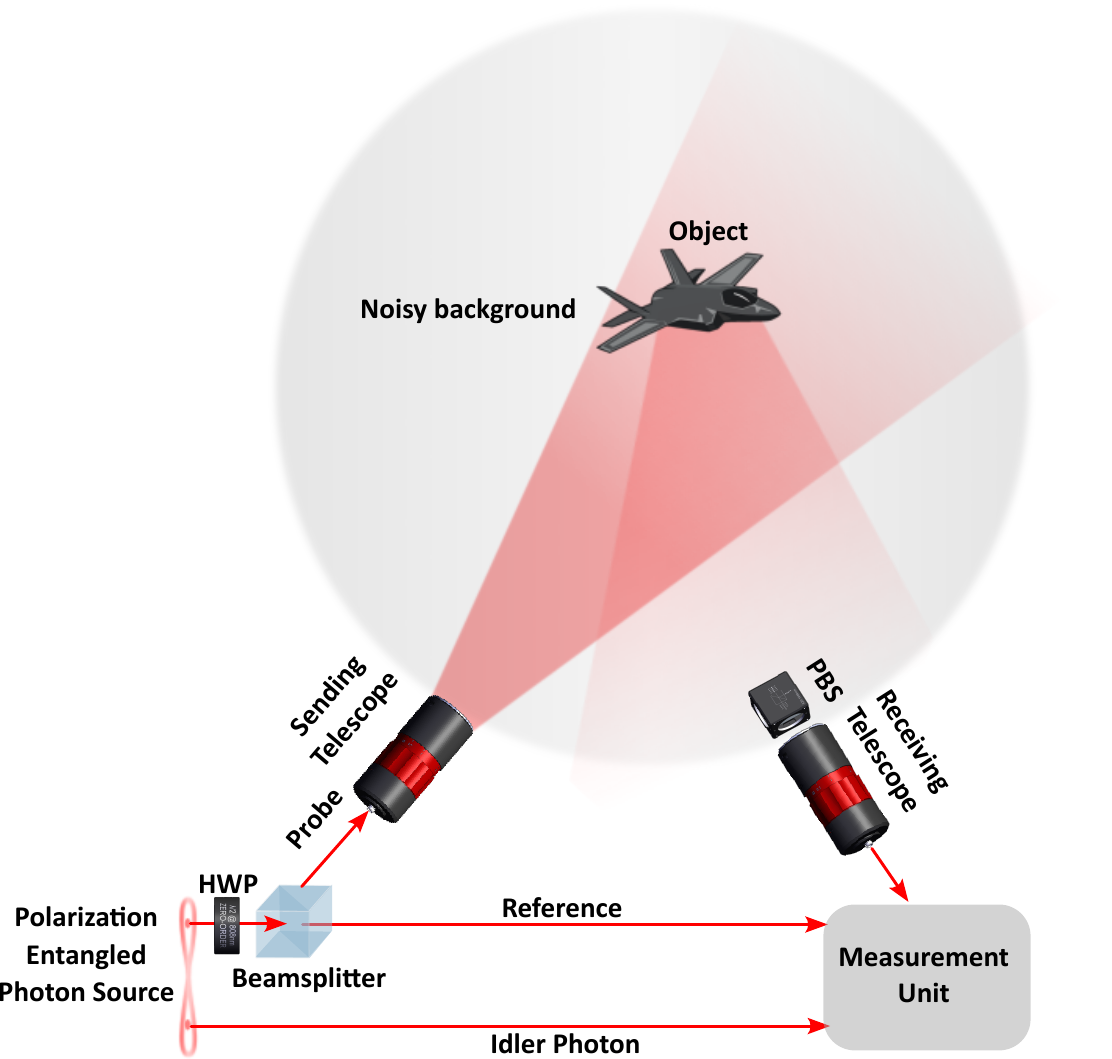}
\caption{Polarization-entanglement based quantum illumination protocol: One photon of each polarization-entangled pair (the idler) is sent directly to the measurement unit, while the other photon (the signal) is probabilistically routed into either the probe or the reference paths via a non-polarizing beam splitter\,(BS). Probe photons are transmitted through a sending telescope towards an object located in a noisy environment. The reflected photons from the object, together with noise, are collected by a receiving telescope after polarization filtering using a polarizing beam splitter (PBS) and are sent to the measurement unit. Photons traveling along the reference path bypass the object and are sent directly to the measurement unit, where both probe-idler and reference-idler correlations are independently analyzed via coincidence detection and correlation (CHSH value) measurements.}
\label{fig: scheme}
\vspace{2mm}
\end{figure}

\noindent
Here, we report a robust QI protocol employing polarization-entangled photon pairs distributed across three paths. The scheme is designed such that the generated entangled photon pairs are encoded in different polarization basis states optimized for the CHSH value measurement before transmitting towards the object. At the receiving end, polarization preserving photons are filtered out to enable $S$-value determination for the photons in two different pairs of paths, making the scheme robust for reliable object detection and ranging in a noisy and lossy environment. The scheme is experimentally validated in a realistic, lossy environment demonstrating object detection over distances approaching $500$ meters, corresponding to an $1$ km round-trip path. As illustrated in Fig.\,\ref{fig: scheme}, polarization-entangled photon pairs are employed for the detection of a distant object embedded within a noisy background. For each pair, the idler photon is routed directly to the measurement unit, while the path for the signal photon is divided into two - labeled as \textit{reference} and \textit{probe} paths. The photons propagating along the reference path are directly sent towards the measurement unit,  whereas the photons in the probe path illuminates the distant object and those reflected back from the object are collected and directed to the measurement unit. Joint correlation measurements performed between the idler photons and the photons in the probe or the reference paths enable independent evaluation of the CHSH Bell parameter for the probe-idler and reference-idler subsystems. The impact of the atmospheric channel on the shared entanglement can be quantitatively inferred by comparing the $S$-values obtained for the probe-idler pairs against those measured for the reference-idler pairs. In addition, the recorded time delay between the arrivals of idler and probe photons provides the range information. A measured CHSH-value exceeding $2$ in the probe-idler correlations confirms the presence of an object, while the temporal information of photon arrivals enables ranging - thereby, demonstrating simultaneous object detection and ranging with quantum advantage. We report CHSH violations with $S > 2.6$ even when a few tens of probe photons are detected upon reflection from the object.\\

%=======================================================================

\noindent
\textit{Description of the protocol:--} The scheme starts with the preparation of a two-qubit polarization entangled state as follows,
\begin{align}
|\Psi_{0}\rangle=\frac{1}{\sqrt{2}}\big(|H\rangle_s|V\rangle_i+|V\rangle_s|H\rangle_i \big),
\end{align}
where subscripts `$s$' and `$i$' denote the signal and idler photons generated from a SPDC source and $|H\rangle$ and $|V\rangle$ respectively represent the horizontal and vertical polarizations of the photons. Demonstrating Bell’s violation using the CHSH Bell parameter requires polarization projection measurements across different basis\ \cite{kwait1995, kanad2024}.
Probabilities of finding the photons in different polarization states in a particular basis can be determined from the coincidence measurement outcomes for a particular setting $(\alpha, \beta)$, where the parameters $\alpha$ and $\beta$ are the angles that control the polarizations of the signal and the idler photons, respectively. The correlation parameter $E(\alpha,\beta)$ is computed as
\begin{align}
    E(\alpha,\beta) = P_{HH} + P_{VV} - P_{HV} - P_{VH},
\end{align}
where $P_{kk'}$ denotes the probability of the photons being in the two-qubit state $|k\rangle_{s}|k'\rangle_{i}$ with $k,k'\in \{H,V\}$. Further, determination of the CHSH parameter (the $S$-value) requires four such settings and is given by,
\begin{align}
    S=|E(\alpha,\beta)-E(\alpha,\beta')+E(\alpha',\beta)+E(\alpha',\beta')|.
    \label{eq: S_pol}
\end{align}
Experimentally the correlation parameter $E(\alpha,\beta)$ is calculated using the coincidence counts, as follows 
\begin{align}
    E(\alpha,\beta) = \frac{N(\alpha,\beta)+N(\alpha^\perp,\beta^\perp)-N(\alpha^\perp,\beta)-N(\alpha,\beta^\perp)}{N(\alpha,\beta)+N(\alpha^\perp,\beta^\perp)+N(\alpha^\perp,\beta)+N(\alpha,\beta^\perp)},
    \label{eq: E_expt}
\end{align}
where $\alpha^\perp$ (or $\beta^\perp$) denote the angle corresponding to the polarization setting orthogonal to $\alpha$ (or $\beta$) for the signal (or idler) photons and $N(\alpha,\beta)$ represents the recorded coincidence counts for the polarization rotation angles $\alpha$ and $\beta$ in the signal and idler channels, respectively.\\

\noindent
As depicted in Fig.\ref{fig: scheme}, the idler photons and the signal photons propagating along the reference path are routed directly to the measurement unit, while the probe photons travel through a lossy free-space channel to illuminate an object (of reflectivity $\mathcal{R}$, say) embedded in a noisy background. The photons reflected from the object are collected back and subsequently directed to the measurement unit for joint detection. Therefore, considering the effects of the object as well as the environment, the probe photons experience an overall attenuation characterized by a factor
\begin{align}
    \eta(L) = \mathcal{N}_{p} \mathcal{R} \exp(-aL)
    \label{eqn: attenuation}
\end{align}
where, $\mathcal{N}_{p}$ is the overall system transmissivity that incorporates both the fraction of transmitted probe photons incident on the object and the fraction of reflected photons successfully collected and coupled into the receiving unit; $L$ represents the roundtrip propagation distance of the probe photons, and $a$ denotes the effective attenuation coefficient of the atmospheric channel accounting for absorption, scattering and turbulence induced losses encountered during propagation. The factor $\mathcal{N}_{p}$ depends on the beam waist, beam divergence, losses in the transmitting and receiving optics, receiver collection efficiency, geometry of the object, its orientation, etc.  \\

\noindent
The half-wave plate (HWP) at angle $\theta_{s}$ in the signal path, as shown in Fig.\,\ref{fig: scheme}, tunes the polarization of the probe and the reference photons simultaneously. Similarly, another HWP at angle $\theta_{i}$ in the idler path, within the measurement unit, controls the idler polarization for the correlation measurement. The measurement settings $\theta_{s} =\frac{\alpha}{2}$ and $\theta_{i}=\frac{\beta}{2}$ rotates the signal and idler polarizations to $\alpha$ and $\beta$ with respect to the horizontal and the correlation measurement $E(\alpha, \beta)$ is performed by coincidence detections after the PBSs in both the paths. The final state at the setting $(\theta_{s}, \theta_{i}) = (\frac{\alpha}{2} , \frac{\beta}{2})$ before the projective measurements are made, can be expressed as 
\begin{align}
    \begin{split}
        |\Psi_{f}(\alpha, \beta)\rangle =& \frac{\sqrt{\eta(L)}}{2} \left[ \sin(\alpha+\beta) \left(|H\rangle_{p}|H\rangle_{i} - |V\rangle_{p}|V\rangle_{i}\right) \right.  \\
        & \left. -\cos(\alpha+\beta) \left(|H\rangle_{p}|V\rangle_{i} + |V\rangle_{p}|H\rangle_{i} \right) \right] \\
        & + \frac{1}{2} \left[ \sin(\alpha+\beta) \left(  |H\rangle_{r}|H\rangle_{i} - |V\rangle_{r}|V\rangle_{i}\right) \right. \\
        & \left. -\cos(\alpha+\beta) \left( |H\rangle_{r}|V\rangle_{i} + |V\rangle_{r}|H\rangle_{i} \right) \right] 
    \end{split}
\end{align}

\noindent
where, the subscripts `$p$' and `$r$', respectively, represent the photons in probe and reference paths. The attenuation affects the coincidence counts, but the correlation parameter - being a ratio of the linear combinations of coincidence counts - should ideally remain unaltered, as the attenuation factors cancel out. Therefore, the correlation parameter evaluated for the reference-idler and probe-idler pairs take the same functional form,
\begin{align}
    E_{r}(\alpha, \beta) = E_{p}(\alpha, \beta)= -\cos2(\alpha+\beta). 
\end{align}
Using the correlation parameters evaluated for different polarization settings $(\theta_{s}, \theta_{i})$, the $S$-values can be determined for various object distances. However, in real experimental scenario when the coincidence counts become extremely low, even minor fluctuations in the measured coincidence counts could lead to a noticeable reduction in the $S$-value. To simultaneously monitor the source entanglement through reference-idler correlations while probing the presence of an object through probe-idler correlations, we evaluate two sets of $S$-values. Therefore, the $S$-value detection for the reference-idler pair serves as a useful resource for long distance QI experiments. \\

%==========================================================================================
%%%%% EXPERIMENT %%%%%

\begin{figure*}[t!]
\centering
\includegraphics[width = 0.98\linewidth]{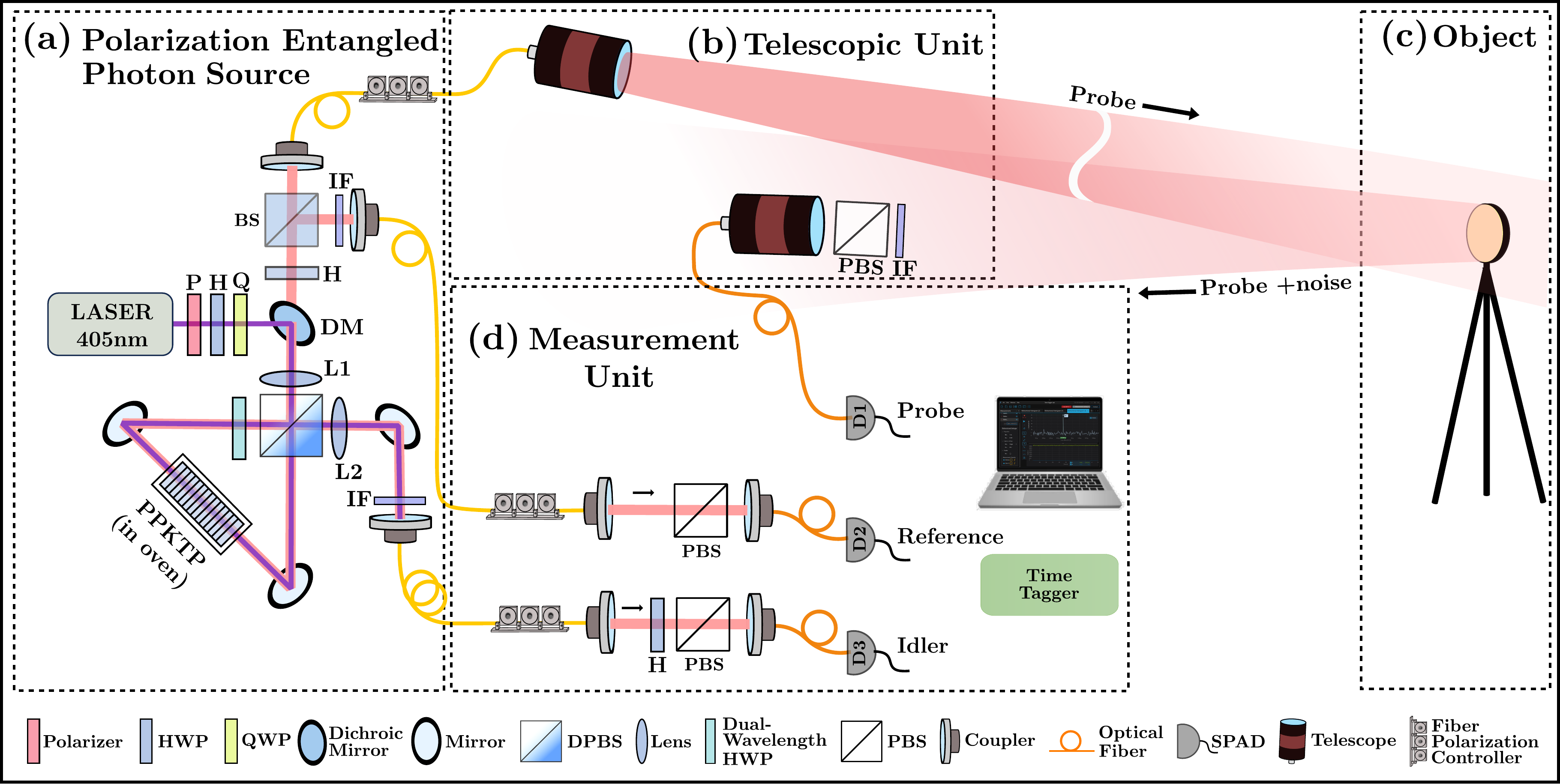}
\caption {Experimental setup for the quantum illumination based object detection. (a) Source: A type-II PPKTP crystal kept within a  Sagnac configuration is pumped using a 405 nm laser that generates polarization entangled photon pairs at 808.049 nm. (b) Telescopic unit: A sending telescope in the probe path transmits the signal photons towards a distant object and a receiving telescope collects the signal reflected back from the object. (c) Object: The object is placed in a noisy environment at a certain distance from the source. (d) Measurement unit: Correlation measurement between idler photons and signal photons from either the probe or the reference path is performed to evaluate the $S$-values.}
\label{fig: expt_setup}
\end{figure*}

\noindent
\textit{Experimental realization:--} The schematic of the experimental setup for the proposed QI scheme is shown in Fig.\,\ref{fig: expt_setup}. The polarization entangled photon pairs are generated via SPDC using a $10$ mm long periodically poled potassium titanyl phosphate (PPKTP) non-linear crystal [Raicol] with poling period $\Lambda = 10\,\mu$m and aperture size of $1\times2 ~\text{mm}^2$, placed inside a common path Sagnac interferometer\,(SI) configuration\,\cite{kim2006}. The crystal, cut for collinear type-II SPDC, is pumped using a $405$ nm laser [Coherent OBIS LX SF] with maximum output power of $40$ mW, having spectral linewidth less than $0.3$ pm. The polarizer ensures that the incoming pump beam is linearly polarized, whereas the HWP and the quarter-wave plate\,(QWP) transform the polarization of the pump to provide power balance (in the two paths of the interferometer) and phase control, required for efficient entangled photon generation from the bidirectionally pumped Sagnac down-conversion source. The dichroic mirror\,(DM) directs the pump beam into the SI, built using two mirrors and a dual-wavelength polarizing beam splitter (DPBS). The lens L1\,($f=200$ mm) focuses the pump beam onto the center of the crystal from both the sides. \\

\begin{figure}[b!]
\vspace{-5mm}
\centering
\includegraphics[width=0.48\textwidth]{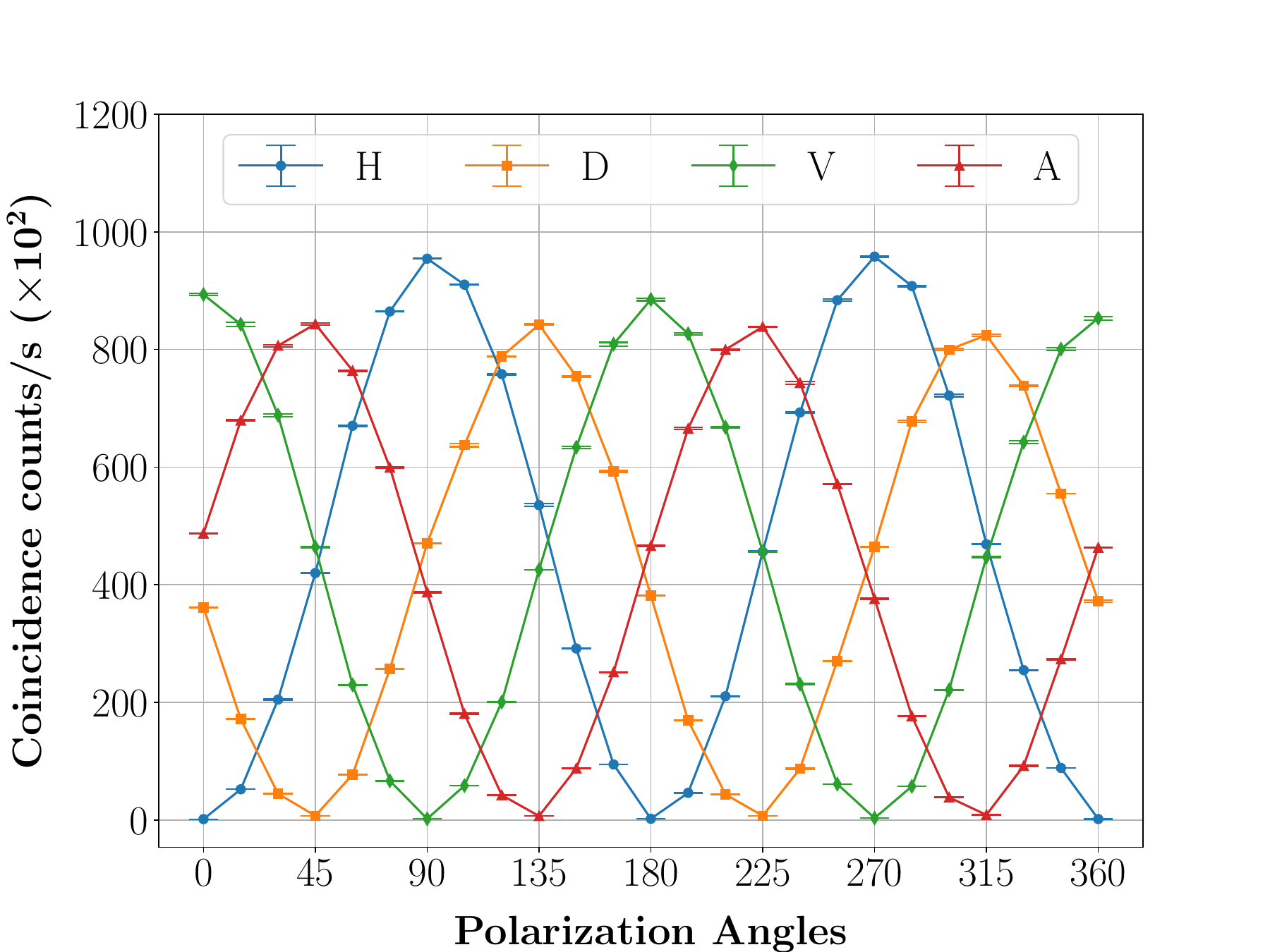}
\caption {Experimentally obtained visibility of the polarization-entangled state. The measured visibilities are $99.5\%$ in H/V basis (blue, green), and $98.4\%$ in A/D basis (orange, red).}
\label{fig: visibility}
\end{figure}

\noindent
The PPKTP crystal is placed inside a temperature-controlled oven and maintained at a temperature of $30^{\circ}$C, that achieves degenerate photon pair generation at $808.049$ nm. The dual wavelength half-wave plate\,(DHWP) oriented at $45^{\circ}$ inside the SI rotates the $V$-polarized pump to $H$-polarized, ensuring proper phase matching in the non-linear crystal. Lenses L1 and L2 ($f=200$ mm) collimate the signal and idler photon beams at the two output ports of the SI. A bandpass interference filter (IF) at $810\pm10$ nm is placed in front of the collection optics to selectively transmit the SPDC photons while suppressing the residual pump. Through polarization control of the pump beam and proper alignment of SI, a single photon source of high flux\,($\sim 45,000$ pairs/s/mW) and high heralding efficiency\,($\sim38\%$) is achieved. The visibility of the polarization-entangled photon state is obtained to be $99.5\%$ in H/V basis and $98.4\%$ in A/D basis, as can be seen in Fig.\,\ref{fig: visibility}. A maximum CHSH value $S=2.802\pm0.002$ is achieved for the prepared state $\textstyle |\Psi^{+}\rangle = (|H\rangle_{s}|V\rangle_{i}+|V\rangle_{s}|H\rangle_{i})/\sqrt{2}$. \\

\noindent
After characterizing the source, the idler photons are sent directly to the measurement unit which consists of a HWP and PBS, while the signal photons are directed through a HWP to a 50:50 BS that splits the signal into reference and probe paths. The reference photons are directly routed to the measurement unit, whereas the probe photons are routed to the measurement unit via an object embedded in a noisy environment.
Note that the HWP, that encodes the signal photons into the measurement bases, is kept before the BS which in turn enables simultaneous CHSH-Bell parameter measurements for probe-idler and reference-idler pairs, using independent PBSs in each arm. This real-time characterization of the source ensures that any reduction in probe-idler $S$-value will be solely due to the object reflectivity and environmental noise. \\

\noindent
Probe photons are transmitted toward the object through a fiber collimator [C80APC-B, Thorlabs] of 2-inch aperture size, referred to as the sending telescope. An identical fiber collimator at the receiving end collects the photons reflected back from the object and directs them to the measurement unit. Considering the limited collection efficiency of single-mode fiber, a multi-mode fiber is used at the receiver, which scrambles the polarization information. Consequently, polarization analysis as part of the correlation measurement for the probe photons, is performed by a PBS located immediately before the receiving telescope. The objects used in the experiment include a gold-coated mirror of diameter 2-inch with approximately $96\%$ front-surface reflectivity and a similar 2-inch diameter circular scattering object of unknown reflectivity. Each object is mounted on a tripod and positioned at distances ranging from 50 m to 500 m from the source to study the effect of free-space propagation length on signal integrity. After each distance adjustment, precise alignment of the sending and receiving telescopes with the object are performed using a back-tracing laser.\\
 
\noindent
Three single photon counting modules\,(SPCM) $\text{D}_{j=1:3}$ [SPCM-800-44-FC detectors, Excelitas] are placed respectively at the transmitting ports of the PBSs in probe, reference and idler paths. The detectors collect the photons after the projective measurements on different polarization basis states and the correlation measurement is performed using a time correlated single photon counter [Time Tagger Ultra, Swabian Instruments]. To simultaneously evaluate the correlation parameter (Eq.\,(\ref{eq: E_expt})) for the probe-idler and reference-idler pairs, coincidence counts corresponding to different measurement outcomes are recorded between detector pairs D$_1$-D$_3$ and D$_2$-D$_3$,  respectively. The associated probabilities $P_{kk'}$, obtained for various combinations of HWP angle settings, are used to compute the CHSH Bell parameter $S$, as defined in Eq.\,(\ref{eq: S_pol}), which quantifies the degree of non-classical correlation for both probe-idler and reference-idler subsystems. Furthermore, the object distance from the source is inferred from the timing information of the recorded coincidence peak, thus facilitating ranging measurements.\\

%============================================================================
%%%%% RESULT and DISCUSSION %%%%%

\noindent
\textit{Results and Discussion:--} Degenerate pairs of polarization-entangled photons are generated with the count rates being about $332kHz\pm 31kHz$. After 50:50 BS, the reference path is directed to the measurement unit with the recorded single photon count rate being $153kHz\pm 11kHz$. This rate is expected to be the same for the photons in the probe path prior to coupling into the sending telescope. The photons in the probe path traverse a free-space atmospheric channel between the sending and receiving telescopes, and encounter losses from multiple origins, including coupling inefficiency, beam divergence, diffraction losses, atmospheric absorption, scattering, object reflectivity, pointing fluctuations, dispersion, etc. Based on the number of photons launched into the probe path, as inferred from the reference photon counts, we have theoretically estimated the expected number of photons at the receiver end incorporating the losses that affect the most.\\

\noindent
The photons in the probe path are coupled into the sending telescope of focal length $80$ mm and clear aperture $42.5$ mm after a fiber polarization controller (FPC) that maintains the polarization of the photons in the probe path. The mode field diameter (MFD) of the single mode fiber (SMF) used here is $5\pm 0.5\,\mu$m, which introduces a coupling loss of about $25-26\%$ as well as a beam divergence of about $0.0035^{\circ}$ at the output of the telescope. The divergence is calculated using the expression,
\begin{align}
\theta \approx \left(\frac{[MFD]}{f}\right)\left(\frac{180}{\pi}\right),
\label{eq: divergence}
\end{align}
where $f$ is the focal length of the collimator. This expression provides a theoretical estimate of the full angle beam divergence at the collimator output, assuming a Gaussian intensity profile of the light emerging from the fiber. 
The beam width at the object distance (say, $d$ from the front focal plane of the collimator) can be determined by evaluating the $1/e^{2}$ beam diameter of the emergent beam at the front focal plane of the collimator and using the beam divergence ($\theta$) as follows,
\begin{align}
  D \approx 4\lambda\left(\frac{f}{\pi[MFD]}\right) + 2d\tan(\frac{\theta}{2}) ~,
 \label{eq:beamdiameter}
\end{align}
where, $\lambda$ is the wavelength of the signal photons in the probe path. \\

\noindent
The calculated beam diameters at the object plane at distances of $50$ m, $100$ m, $150$ m, $200$ m, $300$ m, $400$ m, and $500$ m are obtained as $19.55$ mm, $22.61$ mm, $25.66$ mm, $28.72$ mm, $34.83$ mm, $40.93$ mm, and $47.04$ mm, respectively. Both the objects used in the experiment each have a diameter of $2$-inch ($=50.8$ mm), which ideally accommodate the entire incident beam. However, in practice, atmospheric turbulence, beam pointing fluctuation and alignment imperfections can cause a portion of the beam to fall outside of the object, resulting in a loss of a fraction of the probe photons. 
The beam incident on the object gets reflected depending on the object reflectivity $\mathcal{R}$ and it further diverges during its transit from the object to the receiving unit. The receiving unit consists of a $1$-inch PBS with clear aperture $20.32$ mm, followed by a telescope with a clear aperture of $42$ mm, that collects only a portion of the reflected beam falling upon it. \\

\noindent
Apart from the losses due to the divergence and coupling inefficiency, the photons in the probe path undergo attenuation due to atmospheric conditions during propagation through free space. As the beam travels a round trip distance of $L$ in between the two telescopes, the photon number $N_0$ reduces to $N$ as follows
\begin{align}
    N = N_{0}\exp(-a L),
    \label{eq: beer-lambert}
\end{align}
where $a$ is the attenuation coefficient of the atmosphere that includes scattering and absorption in the medium. The atmospheric attenuation is considered for an altitude of $2,785$ ft above sea level, for the location where the experiment was carried out. We estimate the number of probe photons arriving at the measurement unit by taking into account all these losses. \\

\begin{figure}[h]
\centering
\includegraphics[width=0.48\textwidth]{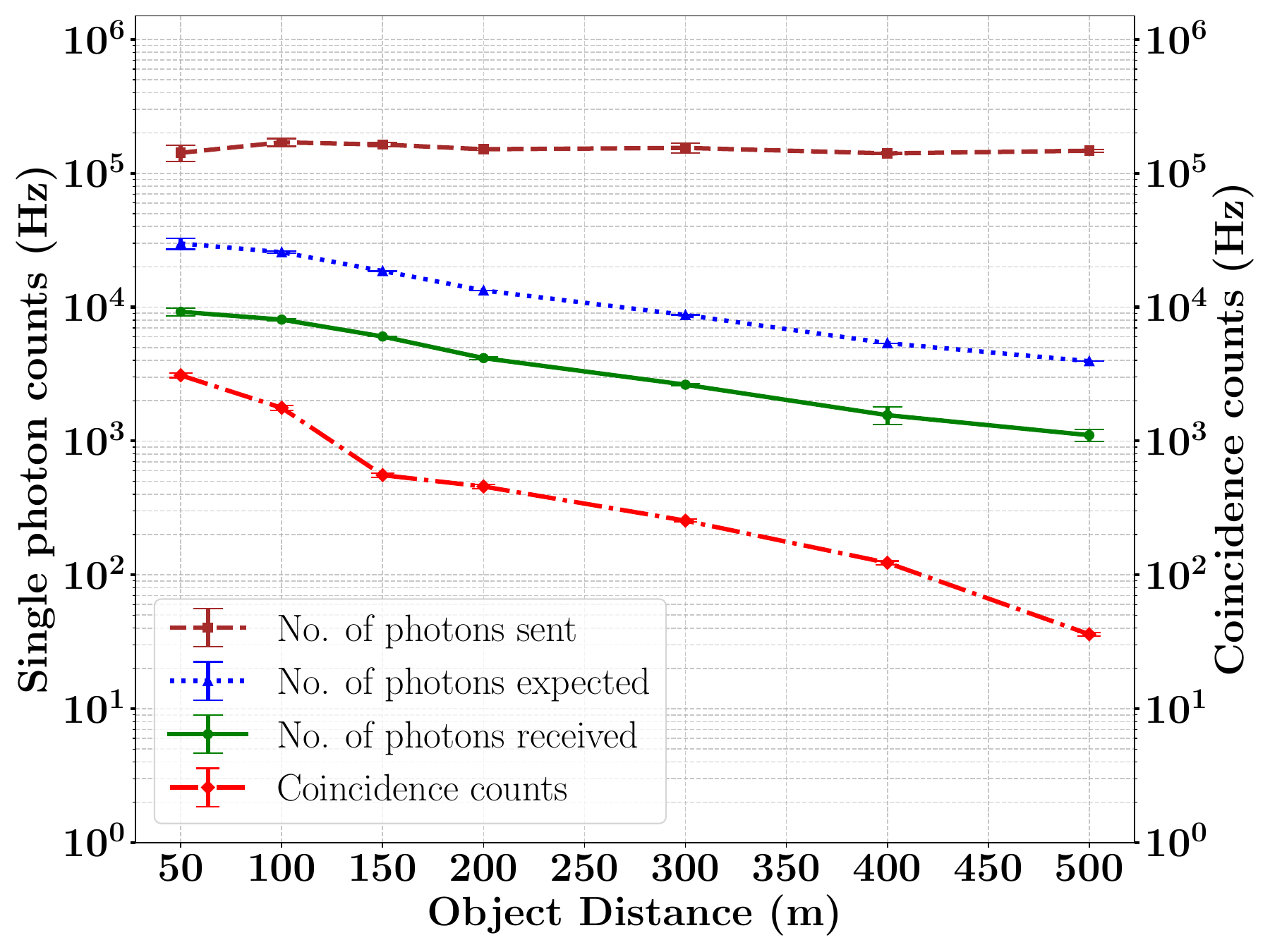}
\caption {Photon counts per second (log scale) as a function of object distance for an object with $\sim96\%$ reflectivity : The brown dashed curve shows the probe photon transmission rate toward the object, while the green solid curve represents the photon collection rate after reflection from the object. The dotted blue curve indicates the theoretically expected collected photon rate after accounting for all system losses. The red dash-dotted curve corresponds to probe-idler coincidence count rate per second as a function of object distance.}
\label{fig:counts_96}
\end{figure}

\noindent
In Fig.\,\ref{fig:counts_96}, the number of photons transmitted per second as probe for object detection, along with the corresponding number of photons expected to be received upon reflection from an object with approximately $96\%$ reflectivity taking into account various possible losses at different distances are plotted. It also presents the experimentally collected photon count rates at various object distances, along with the corresponding probe-idler coincidence count rates recorded with a bin width of $1$ ns. The plot demonstrates that, despite the limited number of probe photons received, the coincidence counts distinctly confirm the presence of the object and serve as a resource to calculate CHSH value using the coincidence counts in different basis for the probe. \\

\begin{figure}[h]
\centering
\includegraphics[width=0.48\textwidth]{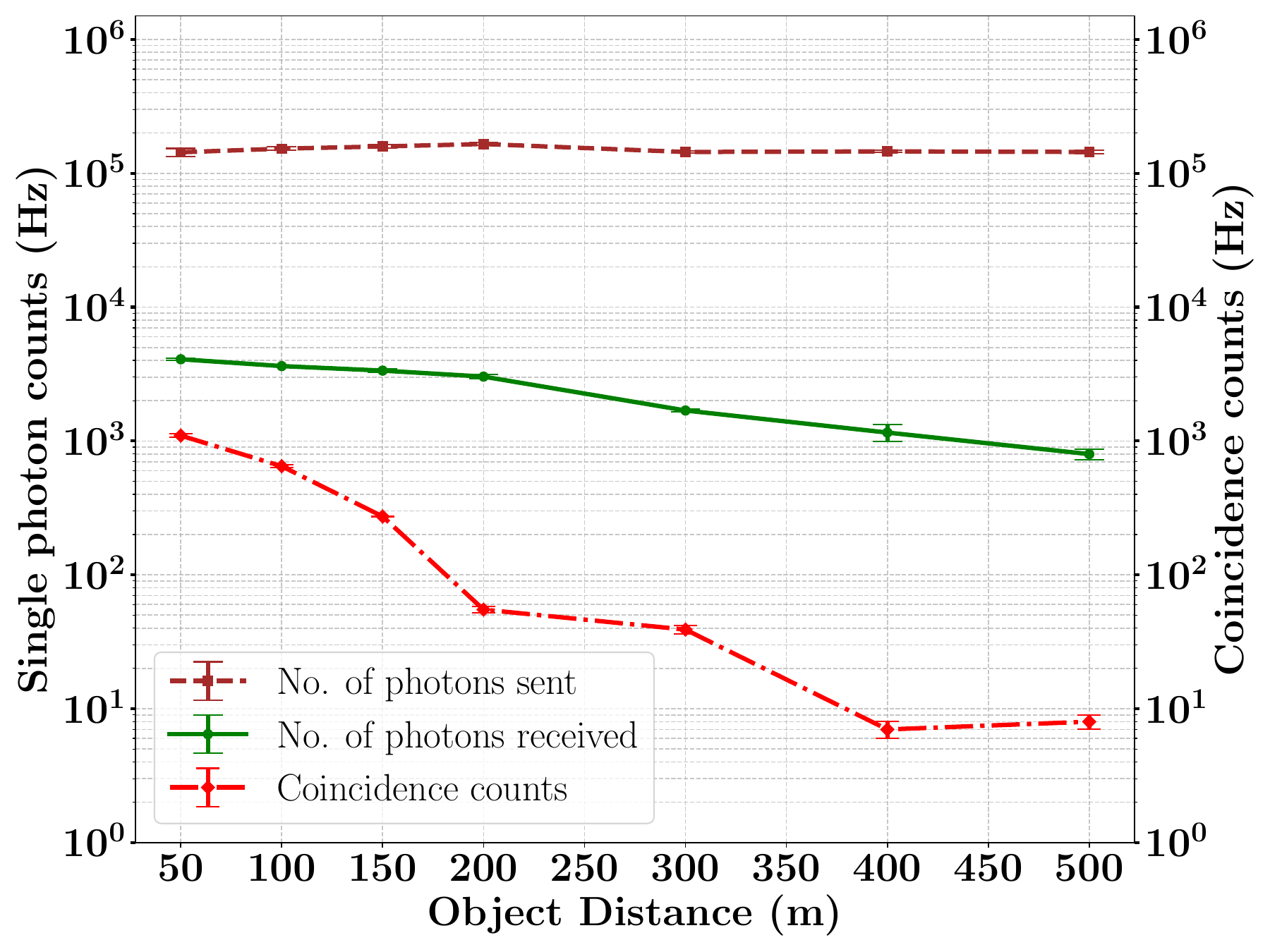}
\caption{Photon counts per second (log scale) as a function of object distance for a scattering object with unknown reflectivity: The brown dashed line shows the probe photon transmission rate toward the object, while the green curve represents the photon collection rate after reflection from the object. The red dash-dotted line corresponds to probe-idler coincidence count rate as a function of object distance.}
\label{fig:counts_unknown}
\end{figure}

\noindent
Similarly, a scattering object of unknown reflectivity of $2$-inch diameter is positioned at distances ranging from 50 m to 500 m. Probe photons are directed towards the object through the sending telescope, and the receiving telescope collects a fraction of the photons that are reflected back from the object. At the measurement unit, joint correlation detection is performed between the collected photons and the idler photons. Figure\,\ref{fig:counts_unknown} shows the probe photon transmission rate toward the object, along with the photon collection rate at the receiving end and the associated coincidence rates. By comparing the received photon count rates (green curves) and the corresponding coincidence rates (red curves) from Fig.\,\ref{fig:counts_96} and Fig.\,\ref{fig:counts_unknown}, we infer that the reflectivity of this unknown scattering object is lower than that of the mirror. Therefore, our scheme can be employed to estimate the unknown reflectivity of an object by combining the measured signal strength with the ranging information, provided we collect all the reflected photons. \\

\begin{figure}[t!]
\centering
\includegraphics[width=0.49\textwidth]{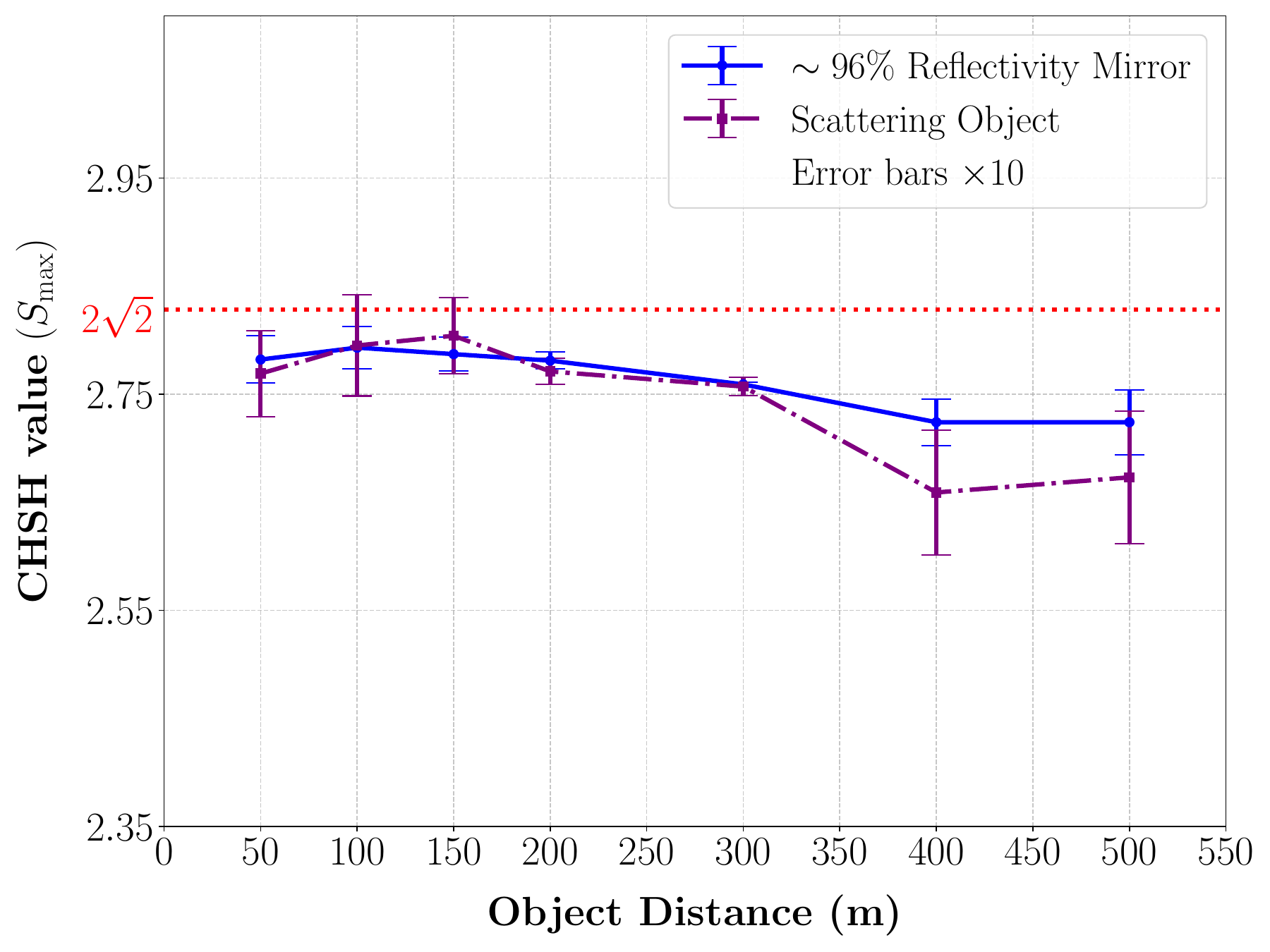}
\caption{Variation of the CHSH $S$-value with object distances. }
\label{fig: svalue}
\end{figure}

\noindent
Figure\,\ref{fig: svalue} shows the experimentally obtained CHSH $S$-values for different distances of the objects - both with known and unknown reflectivities. The plot shows that the measured $S$-values exceed $2.6$ at all object positions, confirming that the detected photons are indeed the probe photons used to illuminate the object, despite the presence of background noise. This shows the robustness of the proposed scheme for long-range object detection.\\

\noindent
The expected photon collection rate in Fig.\,\ref{fig:counts_96} and Fig.\,\ref{fig:counts_unknown} are higher compared to the actual photon collection rate. The theoretical analysis done for expected photon collection rate assume perfect beam alignment, i.e., the probe beam is considered to be perfectly centered on the object as well as on the collection optics. In practice, however, the beam may not be perfectly centered, leading to additional photon loss beyond that arising from atmospheric turbulence, pointing-fluctuations and optical absorption - thereby, reducing the received photon counts.  At longer distances, detector dark counts begin to dominate over the actual probe signal returned after reflection and the coincidence rates become significantly low. This results in a reduction in the measured $S$-values accompanied by larger error bars at those positions as can be seen from Fig.\,\ref{fig: svalue}.\\

\begin{figure}[h]
\centering
\includegraphics[width=0.49\textwidth]{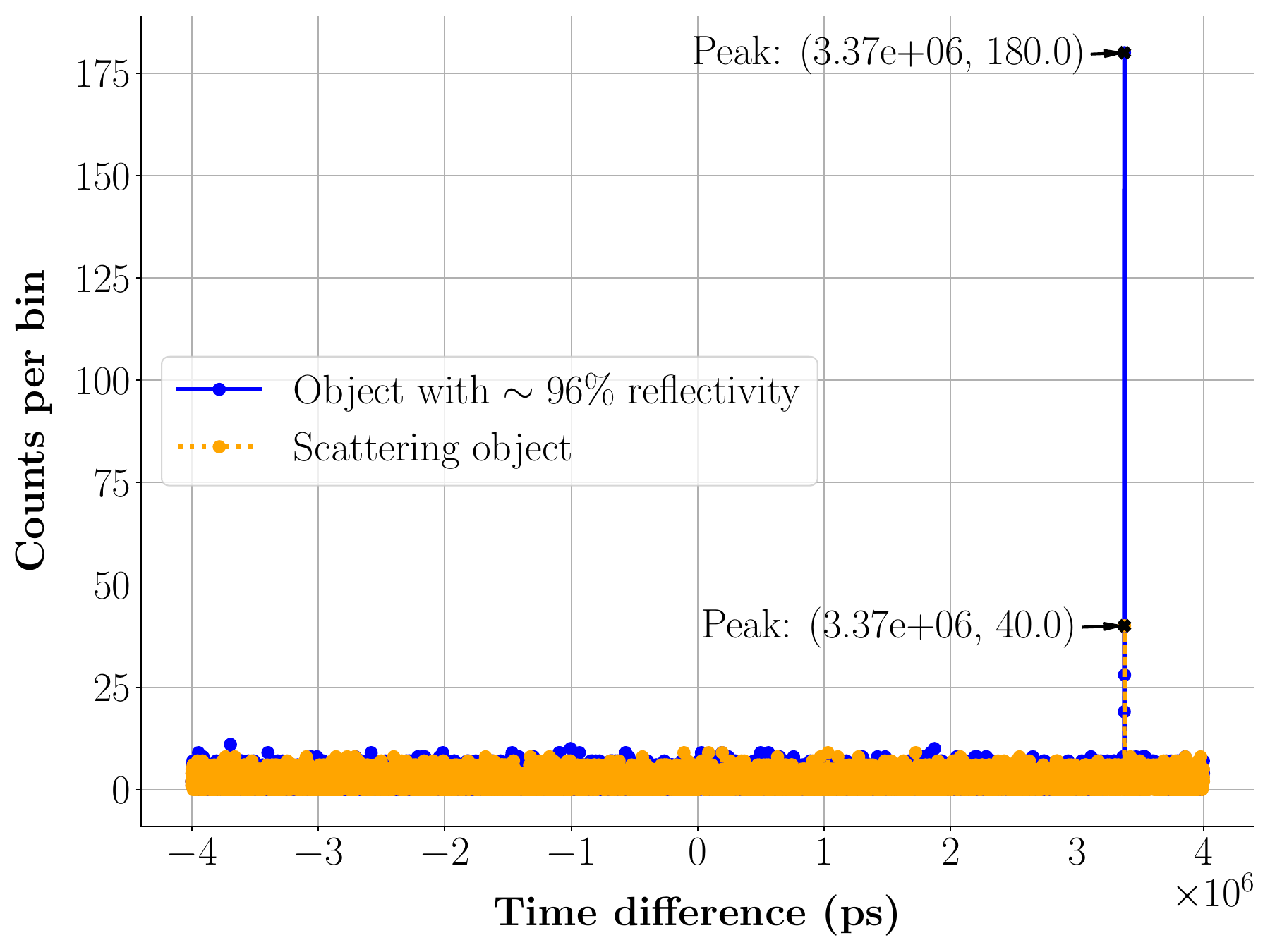}
\caption{Coincidence counts recorded for $5$s as function of time delay between the probe and idler photons with a bin width of $1$ ns for the object distances of $500$ meters.}
\label{fig: coin}
\end{figure}

\noindent
The coincidence peaks captured when the objects are placed at a distance of $500$ m away from the source is shown in Fig.\,\ref{fig: coin}, where the coincidence is measured for $5$ s with a binwidth of $1$ ns. The plot shows $180$ coincidences and $40$ coincidences, respectively received for the object with $\sim96\%$ reflectivity and the scattering object with unknown reflectivity at a time delay of $3.37\times 10^6$ ps. This delay in arrival of the probe photon with respect to the idler photon directly signifies the roundtrip propagation distance of $\sim1000$ m for the probe photons, detecting the object located at $\sim 500$m. Therefore, our scheme enables accurate ranging of a distant object from the timing information of the recorded coincidence peak between probe and idler. \\

\noindent
During field experiments, the entanglement measure between reference and idler was continuously monitored to verify the quality of the source and to proportionally maximize the coincidence counts for the probe and idler paths. Any coincidence events between probe and reference was treated as background noise and the coincidence readings between probe and idler at those particular time stamps were discarded. However, the experiment was carried out at night in a remote location with the source placed in a closed cabin, the object in an abandoned runway and spectral filters before the collection optics. Due to this the background photons were observed to be minimal throughout the experiment.\\

%================================================================================
%%%%% CONCLUSION %%%%%

\noindent
\textit{Conclusion:--}
We have experimentally demonstrated a novel QI protocol based on polarization-entangled photon pairs, using violations of the CHSH inequality as a quantitative measure of non-classical correlations, capable of long-distance object detection and ranging. The performance of the proposed scheme is experimentally validated in a lossy free-space channel over round-trip propagation distances approaching $1$ km. Remarkably, strong quantum correlations were observed even when the return probability of the probe photons was as low as few tens of photons per second, highlighting the robustness of polarization entanglement over longer distances. The current experimental performance is primarily limited by practical factors, including the finite $2$-inch aperture of the receiving telescope, alignment precision, and atmospheric beam wander, all of which reduce the collection efficiency of the returned photons. These constraints are not fundamental and can be addressed by employing receiving telescopes with larger collection areas and incorporating fast steering mirrors with active feedback mechanisms. Thus our scheme can, in principle, operate over substantially longer distances. \\

\vskip 0.05in
\noindent
{\bf  Acknowledgement}\\
We acknowledge the support from the Office of Principal Scientific Advisor to the Government of India, project no. Prn.SA/QSim/2020.
%============================================================
\clearpage

%========================================================

\end{document}